\documentclass[aps,eqsecnum,preprintnumbers]{revtex4}

\begin{document}

\preprint{physics/0302070}
\preprint{SINP/TNP/03-02}
\title{
Reflection and refraction of electromagnetic waves at
the plane boundary between two chiral media}

\author{Jos\'e F. Nieves}
\affiliation{Laboratory of Theoretical Physics, 
Department of Physics, P.O. Box 23343\\
University of Puerto Rico, 
R\'{\i}o Piedras, Puerto Rico 00931-3343}
\author{Palash B. Pal}
\affiliation{Saha Institute of Nuclear Physics, 1/AF Bidhan-Nagar\\
Calcutta 700064, India}
\date{February 2003}

\begin{abstract}

This work is concerned with the propagation of electromagnetic waves
in isotropic chiral media and with the effects produced by a plane
boundary between two such media.  In analogy with the phenomena of
reflection and refraction of plane electromagnetic waves in ordinary
dielectrics, the kinematical and dynamical aspects of these phenomena
are studied, such as the intensity of the various wave components and
the change in the polarization of the wave as it crosses the boundary.
As a prerequisite of this, we show that the plane wave solution must
be written as a suitable superposition of the circularly amplitudes on both
sides of the interface, we elucidate which is the appropriate set of
conditions that the solution must satisfy at the boundary, and we
set down the minimal, and complete, set of equations that must be
solved for the coefficient amplitudes in order to satisfy the
boundary conditions. The equations are solved explicitly 
for some particular cases and configurations (e.g., normal incidence), 
the salient features of those solutions are analyzed in some detail, 
and the general solution to the equations is given as well.

\end{abstract}

\maketitle

%
% section 1
%
\section{Introduction}
\label{sec:introduction}
The problems related to the propagation of electromagnetic waves
in a homogeneous medium, including those of reflection and refraction 
at the interface between two media,
form a traditional set in classical electrodynamics,
and are discussed in any textbook on the subject \cite{jackson}.
A less familiar one is the problem of the propagation of
an electromagnetic wave in a chiral medium.  
In such a medium, the dielectric constant has a different value
depending on the polarization of the wave, which implies
that the two circularly polarized waves propagate with different 
speeds. A chiral medium, also called ``optically active'', can
exist for various reasons. For example, a material medium
can become optically active under the influence of an
external field, such as a magnetic field (the well-known
\emph{Faraday effect} discussed in many plasma physics 
textbooks \cite{faradayeffectbooks}) or a (control) plane
electromagnetic wave \cite{verbiest}.  Other kind of media exhibit the
phenomenon naturally, such as a gas of chiral molecules \cite{charney}, 
or a neutrino gas \cite{np:pip,mnp:nupip} that
contains an unequal number of neutrinos and antineutrinos.  Recently,
man-made composite materials have been produced with similar
properties \cite{nsfpaper}.

Various approaches have been proposed in the literature to study the
electromagnetic properties of such optically active media\cite{condon,
drude, post, charney, varadan, kong1, kong2, krowne, monzon,
hillion3}.  As we showed in Ref.\ \cite{np:thirdconst}, a particularly
simple and general method consists in parametrizing the induced charge
and current in the most general way consistent with general
principles, such as the conservation of the electromagnetic current
and, if applicable, the isotropy and homogeneity of the medium.

However, the formalism developed in Ref.~\cite{np:thirdconst} concerns
a homogeneous medium only.  The problems associated with
electromagnetic wave propagation in inhomogeneous chiral media were
not considered in our earlier work.  An important example of an
inhomogeneous medium is provided by two semi-infinite homogeneous
media, separated by a plane interface.  As is well-known, if an
electromagnetic wave is incident on this interface, part of it is
reflected back into the original medium, and a part is refracted into
the other medium.  The main object of this paper is to extend our
minimal and general parametrization to treat this kind of problem. In
particular, here we study the problem of reflection and refraction at
the interface in detail, for the case in which one or both of the
semi-infinite media on either side of the boundary exhibits the
chirality property.

The paper is organized as follows.  In Section\ \ref{sec:param}, we
summarize the parametrization of the induced charge and current for
a homogeneous chiral media that was introduced 
in Ref.\ \cite{np:thirdconst}. We then
extended it to treat inhomogeneous media, and
we consider in particular the system that consists of 
two semi-infinite homogeneous media
separated by a plane boundary, which is what concerns us in the present 
work.  To set up the stage for the study of wave propagation across
the boundary between two chiral media, 
in Section\ \ref{sec:homo} we review briefly the
salient features of the propagation of a plane wave
in a homogeneous chiral medium.  
Then in Section\ \ref{sec:reflecrefrac}
we set up the problem of reflection
and refraction at the plane interface between two chiral media,
in terms of a set of equations that must be solved for the various
coefficient amplitudes that appear in the suitably decomposed
plane wave solution of the Maxwell equations.
In Section\ \ref{sec:solutions} we solve those equations to obtain the
amplitudes of the reflected as well as refracted waves.
The corresponding solutions for various particular 
cases (including simpler configurations
such as the case of normal incidence), and some of their features,
are analyzed in some detail
before the solution for the general case is given.
Section\ \ref{sec:conc} contains some concluding remarks and outlook.

%
% section 2
%
\section{Parametrization of the induced sources}\label{sec:param}
Here, and in the rest of the paper we deal exclusively with materials
that are isotropic, and we also assume that the field strengths are
such that the material responds linearly. 
We use the Heaviside-Lorentz system of units throughout.

\subsection{Non-dispersive medium}
For the sake of orientation, let us consider first a non-dispersive
medium.  Our argument, advocated in Ref.~\cite{np:thirdconst}, was to
use the Maxwell equations themselves as a guide and write the induced
charge and current densities in the usual form
\begin{eqnarray}
\label{ind}
\rho_{\rm ind} & = & -\vec\nabla\cdot\vec P \,, \nonumber\\
\vec j_{\rm ind} & = & \frac{\partial\vec P}{\partial t} + 
\vec\nabla\times\vec M\,,
\end{eqnarray}
but the relationship between the electric and magnetic polarization
$\vec P$ and $\vec M$ being more general than the usual one. For the
example we are considering, the relation is
\begin{eqnarray}
\label{P&M}
\vec P & = & (\epsilon - 1)\vec E\nonumber\\
\vec M & = & (1 - \mu^{-1})\vec B - \zeta \vec E \,,
\end{eqnarray}
which is consistent with all the general principles stated above.  In
particular, the isotropy of the medium implies that the quantities
$\epsilon,\mu,\zeta$ are scalars rather than tensors, and the
non-dispersive nature of the medium implies those same quantities are
constants independent of the coordinates. As a consequence of the
homogeneous pair of the Maxwell equations,
\begin{eqnarray}
\vec\nabla \cdot \vec B &=& 0 \,, 
\label{divB}\\*
\vec\nabla \times \vec E &=& - {\partial \vec B \over \partial t} \,,
\label{faraday}
\end{eqnarray}
an equivalent parametrization of the induced charge is furnished
by writing 
\begin{eqnarray}
\vec P & = & (\epsilon - 1)\vec E + \zeta \vec B \,,\nonumber\\
\vec M & = & (1 - \mu^{-1})\vec B\,,
\end{eqnarray}
instead of Eq.\ (\ref{P&M}). In any case,
\begin{eqnarray}
\label{inducedparametrization}
\rho_{\rm ind} & = & -(\epsilon - 1)\vec\nabla\cdot\vec E \,, \nonumber\\
\vec j_{\rm ind} & = & (\epsilon - 1)\frac{\partial\vec E}{\partial t} + 
\left(1 - \frac{1}{\mu}\right)\vec\nabla\times\vec B - 
\zeta\vec\nabla\times\vec E\,.
\end{eqnarray}

A question that arises naturally is whether the expressions for
$\vec P$ and $\vec M$ can both contain two terms, one proportional to
$\vec E$ and one to $\vec B$, namely,
\begin{eqnarray}
\label{redundantparam}
\vec P & = & (\epsilon - 1)\vec E + \zeta_1 \vec B \,,\nonumber\\
\vec M & = & (1 - \mu^{-1})\vec B - \zeta_2\vec E\,.
\end{eqnarray}
Indeed, this has been proposed in the literature\footnote{See Ref.\
\cite{np:thirdconst}, and references therein.}.  However, the
discussion above makes it clear that such a parametrization is
redundant, since the expression for the induced sources would again be
given by Eq.\ (\ref{inducedparametrization}), with
\begin{eqnarray}
\zeta = \zeta_1 + \zeta_2 \,,
\end{eqnarray}
and therefore $\zeta$ is the only physically relevant quantity and not
$\zeta_{1,2}$ separately.

\subsection{Dispersive but homogeneous medium}
While these arguments refer to a non-dispersive medium, similar
considerations, and the same conclusions, hold for a dispersive, but
homogeneous medium as well. In a dispersive medium, the relation between
$\vec P$, $\vec M$ and the fields is non-local. Therefore, expressions
like $\epsilon\vec E$ should interpreted in the form
\begin{eqnarray}
\label{nonlocal}
\epsilon\vec E \rightarrow \hat\epsilon\vec E \equiv \int\,d^4x^\prime
\epsilon(x,x')\vec E(x') \,,
\end{eqnarray}
and similarly with the other terms involving $\mu$ and $\zeta$ (in this 
section, we will use the four-dimensional notation $x = (t,\vec x)$ 
in the arguments of the functions for compactness.)
However,
the assumption that the medium is homogeneous implies that the functions
$\epsilon,\mu,\zeta$ depend only on the relative distance
$\vec x - \vec x^\prime$ and, furthermore, 
the causality  principle requires the
time dependence to be on the variable $t - t^\prime$ and not on
$t$ and $t^\prime$ separately. In this case the 
relations such as those in 
Eq.\ (\ref{nonlocal}) become simple algebraic relations in Fourier space
and, as we showed in detail in Ref.\ \cite{np:thirdconst}, the
argument and the parametrization we have used above continue to hold, if every
symbol is taken to be the Fourier transform of the corresponding coordinate
space variable, and if we make the identification
\begin{eqnarray}
\vec\nabla & \rightarrow & i\vec k \,,\nonumber\\
\frac{\partial}{\partial t} & \rightarrow & -i\omega \,.
\end{eqnarray}
Thus in Fourier space, the induced sources are parametrized as
\begin{eqnarray}
\rho_{\rm ind}(\omega,\vec k) & = & -i(\epsilon - 1)\vec k\cdot \vec E\,,
\nonumber\\
\vec j_{\rm ind}(\omega,\vec k) & = &
-i(\epsilon - 1)\omega\vec E + i(1 - \mu^{-1})
\vec k\times\vec B - i\zeta\vec k\times\vec E\,,
\end{eqnarray}
and the corresponding expressions in coordinate space are
\begin{eqnarray}
\label{rhoinddispersive}
\rho_{\rm ind} & = & -\vec\nabla\cdot[(\hat\epsilon - 1)\vec E] \,, \\
\label{jinddispersive}
\vec j_{\rm ind} & = & \frac{\partial}{\partial t}[(\hat\epsilon - 1)\vec E] + 
\vec\nabla\times\left[\left(1 - \frac{1}{\hat\mu}\right)\vec B\right] - 
\vec\nabla\times(\hat\zeta\vec E) \,.
\end{eqnarray}
It should be noted that the last term in Eq.\ (\ref{jinddispersive})
can be written in terms of $\vec B$ using the relation
\begin{eqnarray}
\label{goldenrelation}
\vec\nabla\times(\hat\zeta\vec E) = -\frac{\partial(\hat\zeta\vec B)}
{\partial t} \,,
\end{eqnarray}
which as a consequence of Faraday's law, as can be seen very simply
in Fourier space. In coordinate space, it is proven by noticing that,
for a homogeneous medium as we are considering, we can use manipulations
such as
\begin{eqnarray}
\frac{\partial}{\partial x_i}\int d^4x\, \zeta(x - x^\prime) \vec E(x^\prime)
& = & 
-\int d^4x \left(\frac{\partial}{\partial x^\prime_i}\zeta(x - x^\prime)\right)
 \vec E(x^\prime) \nonumber\\
& = &
\int d^4x\, \zeta(x - x^\prime) 
\frac{\partial}{\partial x^\prime_i}\vec E(x^\prime)\,,
\end{eqnarray}
where we have integrated by parts.

Thus, in
absence of external sources, the Maxwell equations in a homogeneous
chiral medium consist of the pair given in Eqs.\ (\ref{divB}) and
(\ref{faraday}), augmented by
\begin{eqnarray}
\label{gauss}
\vec\nabla\cdot(\hat\epsilon\vec E) & = & 0\,, \\
\label{ampere}
\vec\nabla\times\left(\frac{1}{\hat\mu}\vec B + \hat\zeta\vec E \right) -
{\partial \over \partial t} (\hat\epsilon\vec E) &=& 0\,.
%\label{}
\end{eqnarray}
If the external sources are non-zero, they would
appear on the right hand sides of these two equations.

\subsection{An inhomogeneous medium}
In an inhomogeneous medium, the function $\epsilon(x,x^\prime)$
introduced in the previous section, and similarly $\mu$ and $\zeta$,
depend on the coordinates $\vec x$ and $\vec x^\prime$ separately, and
not just on the relative coordinate $\vec x - \vec x^\prime$.  In this
case, the language of the Fourier transforms is not useful.  Moreover,
whichever way we would like to look at it, the manipulations that led
us to Eq.\ (\ref{goldenrelation}) no longer hold, and therefore in
this case
\begin{eqnarray}
\label{ngoldenrelation}
\vec\nabla\times(\hat\zeta\vec E) \not= -\frac{\partial(\hat\zeta\vec B)}
{\partial t} \,.
\end{eqnarray}
The important implication of this for us is that the question that we
posed ourselves in Eq.\ (\ref{redundantparam}) comes to haunt us
again.  In the present context, it amounts to asking whether instead
of Eqs.\ (\ref{rhoinddispersive}) and (\ref{jinddispersive}) we can
write
\begin{eqnarray}
\label{rhoindinh}
\rho_{\rm ind} & = & -\vec\nabla\cdot[(\hat\epsilon - 1)\vec E] -
\vec\nabla\cdot(\hat\eta \vec B)\,, \\
\label{jindinh}
\vec j_{\rm ind} & = & \frac{\partial}{\partial t}[(\hat\epsilon - 1)\vec E] + 
\vec\nabla\times\left[\left(1 - \frac{1}{\hat\mu}\right)\vec B\right] - 
\vec\nabla\times(\hat\zeta\vec E) + \frac{\partial}{\partial t}
(\eta\vec B)\,.
\end{eqnarray}
The issue now is not whether the two parameters
$\zeta$ and $\eta$ are redundant and whether one
can be absorbed in the other. They are not redundant; the
last two terms in Eq.\ (\ref{jindinh}) are not proportional to each other
in general, and the second term in Eq.\ (\ref{rhoindinh}) is not zero.
Thus, $\zeta$ and $\eta$ have a different and separate physical
meaning, and they describe different physical effects.

The question of which physical
systems are described by this parametrization is one that
we cannot consider here. What we can state is that
there are no general physical principles that exclude that
possibility. Our parametrization is both general and minimal,
subject to the restriction of linearity and isotropy that we have assumed.
However, as we have argued, such systems must necessarily be inhomogeneous.
As interesting as the exploration of these systems would be,
further considerations along these lines in a general way
is outside our scope.

\subsection{Two homogeneous chiral media with an interface}
We finally get to consider the kind of system that concern us in this
work, namely, two semi-infinite homogeneous media, separated by a
plane interface. As we argue below, the proper way to parametrize this
kind of system is by putting $\eta = 0$ in Eqs.\ (\ref{rhoindinh}) and
(\ref{jindinh}).  That is, the Maxwell equations are the same as
stated in Eqs.\ (\ref{gauss}) and (\ref{ampere}).  The argument is
actually subtle, but simple.

Let us take the plane boundary to be the plane defined by $z = 0$.
In the idealized limit of taking the plane boundary
to be infinitesimally thin, then we can represent $\hat\eta$ in the form
\begin{eqnarray}
\label{idealizedeta}
\hat\eta = \eta\theta(-z) + \eta^\prime\theta(z) \,.
\end{eqnarray}
However, on either side of the boundary (that is, for $z \not= 0$),
the function is zero since we are taking both semi-infinite media to
be homogeneous. Therefore, $\hat\eta = 0$ everywhere. 

The exception to this argument arises if the situation is such that
the plane boundary cannot be taken to be infinitesimally thin, but
instead we must take into account the fact that it has some non-zero
thickness $\delta$. Certainly, in the transition from one material
to the other, the medium is inhomogeneous and in that region $\hat\eta$
can be non-zero. If we look at the physical effects that occur
in the bulk of either media at distances, say $|z| \sim L$ with
$L \gg \delta$, then the effects due to the non-zero value of $\hat\eta$
in the transition region are of order $\delta/L$. In other words,
they are surface effects. At points that are close to the plane boundary
($|z| \sim \delta$) such effects could be observable. 
But deep in the bulk of either media
the effects become negligible as $\delta/L\rightarrow 0$,
as it should be since in that case the transition region
can be idealized to be infinitesimally thin and then the representation
given in Eq.\ (\ref{idealizedeta}) is again a valid one.

%
% section 3
%
\section{Wave propagation in an optically active homogeneous medium}
\label{sec:homo}
\subsection{Dispersion relation}
For illustrative purposes, let us consider a steady-state
(monochromatic) electromagnetic wave propagating along a given
direction, which we take to be the $z$ axis. The vector potential is
of the form
\begin{eqnarray}
\label{ex:A}
\vec A(z,t) = e^{-i\omega t}\vec a(z)
\end{eqnarray}
where $\vec a(z)$ must be chosen such that the vector potential
$\vec A(z,t)$ satisfies the Maxwell equations with the
appropriate boundary conditions.  Putting this into 
Eqs.\ (\ref{gauss}) and (\ref{ampere}), it follows that
the wave equation is satisfied only for the circularly polarized waves
whose polarization vectors are defined by 
\begin{eqnarray}
\label{ex:polvectors}
\hat e_{\pm} \equiv \frac{1}{\sqrt{2}}(\hat u_x \pm i u_y) \,,
\end{eqnarray}
where we denote by $\hat u_x$, $\hat u_y$ and $\hat u_z$ the unit vectors 
along the $x$, $y$ and the $z$ directions, respectively.  
Further, these polarization states satisfy the dispersion relations
\begin{eqnarray}
\label{disprel}
\frac{K_\sigma(\omega)}\omega = n_\sigma(\omega,K_\sigma) \,,
\end{eqnarray}
where the refractive index functions $n_\pm$ that appear in 
Eq.\ (\ref{disprel}) are given in terms of the dielectric function
($\epsilon$), magnetic permeability ($\mu$) and activity ($\zeta$)
constants of the medium by\cite{np:thirdconst} 
\begin{eqnarray}
\label{npm}
v_\pm^2 \equiv {1 \over n_\pm^2} = \frac{1}{\epsilon\mu} \pm 
\frac{i\zeta\omega}{\epsilon K} \,.
\end{eqnarray}
While the index of refraction can in general be complex, even without
the $\zeta$ term, in this work we consider only systems for which it
is real; i.e., the medium is non-absorbing. This implies, in
particular, that $\zeta$ is purely imaginary. This, in turn, has
implications on the microscopic properties of the system with respect
to the discrete space-time symmetries, which were considered at
length in Ref.\ \cite{np:thirdconst}.

\subsection{Linearly polarized wave}
For a plane wave moving in the $z$-direction that is linearly
polarized (i.e., it contains equal admixture of the two circular
polarizations), the vector potential is of the form shown in Eq.\
(\ref{ex:A}), where
\begin{eqnarray}
\label{ex:a}
\vec a(z) = A(e^{izK_+}\hat e_+ + e^{izK_-}\hat e_-) \,.
\end{eqnarray}
In writing Eq.\ (\ref{ex:a}) we have implicitly chosen the origin and
orientation of the coordinate system in such a way that, at $z = 0$,
the linear polarization vector of the wave points along $\hat u_x$, or
the $x$-direction.  By simple algebra, and using Eq.\ (\ref{ex:polvectors}),
$\vec A(z,t)$ can be expressed as
\begin{eqnarray}
\label{ex:Afinal}
\vec A(z,t) = A e^{-i\omega t} e^{i\Delta(z)}
\frac{1}{\sqrt{2}}(\hat u_x \cos\theta(z) + \hat u_y\sin\theta(z) ) \,,
\end{eqnarray}
where
\begin{eqnarray}
\label{ex:theta}
\theta(z) & \equiv & \frac{1}{2}(K_{-} - K_{+})z \,,\nonumber\\
\Delta(z) & \equiv & \frac{1}{2}(K_{+} + K_{-})z \,.
\end{eqnarray}
Thus, at a given distance $z = d$, the polarization vector
of the wave points at an angle given by $\theta(d)$ relative
to the $x$-axis, the phenomenon known as optical rotation.

However, it should be noted that for a fixed point $z$,
the direction of the polarization vector is fixed and does not
change with time. This contrasts with what happens for
a circularly polarized wave, e.g.,
\begin{eqnarray}
\label{ex:circular}
\vec A(z,t) = A_{+}e^{-i\omega t}e^{izK_+}\hat e_+ \,.
\end{eqnarray}
In this case the direction of
the electric field is given by the vector
\begin{eqnarray}
\vec\epsilon(z,t) = 
\hat u_x\cos(\omega t - zK_{+}) + \hat u_y\sin(\omega t - zK_{+})\,,
\end{eqnarray}
where, for simplicity, we have taken $A_{+}$ to be real, but similar
considerations apply otherwise. For a fixed point $z$, the
polarization vector rotates in a circle, in the plane perpendicular
to the direction of propagation.

\subsection{Elliptically polarized wave}
\label{subsec:ellipticalwaves}
The obvious generalization of the above special forms is the
superposition
\begin{eqnarray}
\label{ex:elliptical}
\vec A(z,t) = e^{-i\omega t}\left[A_{+}e^{izK_+}\hat e_+ +
A_{-}e^{izK_{-}}\hat e_{-}\right] \,,
\end{eqnarray}
which contains an unequal mixture of both circular polarizations.  A
simple geometric representation of this solution is given by the
following construction.  At the point $z = 0$, the direction of the
electric field is given by the vector
\begin{eqnarray}
\vec \epsilon(z = 0,t) = (A_{+} + A_{-})\hat u_x\cos\omega t +
(A_{+} - A_{-})\hat u_y\sin\omega t \,.
\end{eqnarray}
That is, the polarization vector rotates in an ellipse, with
the axis of the ellipse lying along the $x$ and $y$ directions. 
For an arbitrary point
$z\not= 0$, we write Eq.\ (\ref{ex:elliptical}) in the equivalent form
\begin{eqnarray}
\label{ex:elliptical2}
\vec A(z,t) = e^{-i\omega t}e^{i\Delta(z)}
\left[A_{+} e^{-i\theta(z)}\hat e_{+} + 
A_{-}e^{i\theta(z)}\hat e_{-}\right] \,,
\end{eqnarray}
where $\theta(z)$ and $\Delta(z)$ are defined in 
Eq.\ (\ref{ex:theta}).
The direction of the electric field is easily obtained from this to be
given by the vector
\begin{eqnarray}
\label{poldirectionellip}
\vec\epsilon(z,t) =
(A_{+} + A_{-})\cos(\omega t - \Delta(z))\hat u_{x^\prime} +
(A_{+} - A_{-})\sin(\omega t - \Delta(z))\hat u_{y^\prime} \,,
\end{eqnarray}
where we have defined
\begin{eqnarray}
u_{x^\prime} & = & \cos\theta(z) \hat u_x + \sin\theta(z)\hat u_y
\nonumber\\ 
u_{y^\prime} & = & -\sin\theta(z) \hat u_x + \cos\theta(z)\hat u_y \,.
\end{eqnarray}
Therefore, at a point $z = L$ the polarization is also elliptical,
but in contrast to the standard elliptical polarization phenomena in ordinary
media, the ellipse itself has rotated by the angle $\theta(L)$. In addition,
the polarization angle acquires a negative phase $\Delta(L)$
relative to the $x^\prime$ axis.

In the next sections, we extend these examples to the case in which
the wave crosses the plane interface between two media, each
of which is individually described as we have explained above.
The treatment of that problem must necessarily take into account
the kinematical and dynamical aspects of the reflection and refraction
of the wave at the plane boundary, among other issues.

%
% section 4
%
\section{Reflection and refraction of a circularly polarized wave}
\label{sec:reflecrefrac}
\subsection{Kinematics}
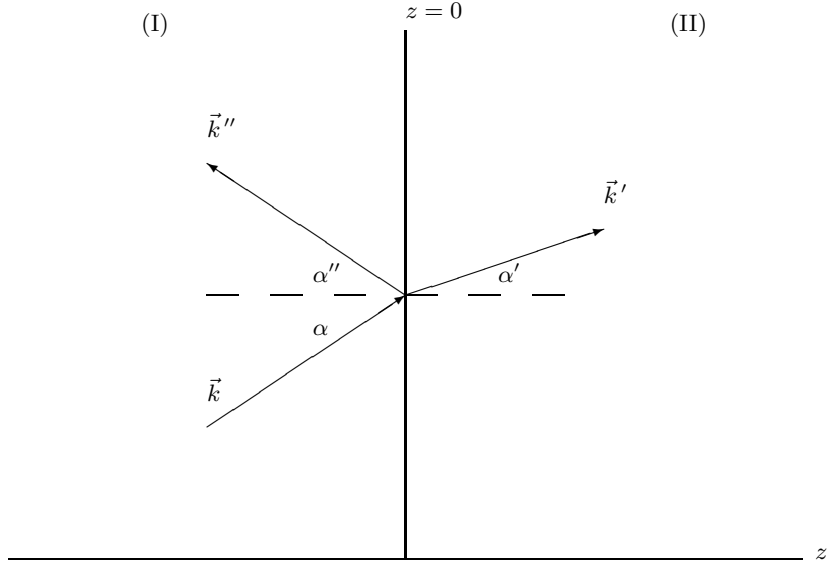
\begin{figure}
\begin{center}
\begin{picture}(300,200)(-150,0)
% region labels
\put(-100,200){(I)}
\put(100,200){(II)}
% axis
\put(-150,0){\line(1,0){300}}
\put(155,0){$z$}
\put(0,205){$z = 0$}
\put(0,0){\line(0,1){200}}
% incident
\put(-75,50){\vector(3,2){75}}
\put(-75,60){$\vec k$}
% reflected
\put(0,100){\vector(-3,2){75}}
\put(-75,160){$\vec k^{\,\prime\prime}$}
% refracted
\put(0,100){\vector(3,1){75}}
\put(75,135){$\vec k^{\,\prime}$}
% angles
\multiput(0,100)(24,0){3}{\line(1,0){12}}
\put(35,105){$\alpha^\prime$}
\multiput(-75,100)(24,0){3}{\line(1,0){12}}
\put(-35,105){$\alpha^{\prime\prime}$}
\put(-35,85){$\alpha$}
\end{picture}

\end{center}
\caption{\label{fig:waves}
Schematic diagram showing a plane electromagnetic wave with wave
vector $\vec k$ incident on a plane boundary at $z = 0$, giving rise
to a refracted and a reflected wave with wave vectors $\vec k'$ and
$\vec k''$, respectively.}
\end{figure}

We consider an electromagnetic wave incident on a plane boundary as
shown schematically in Fig.~\ref{fig:waves}.
%
% fig 
%
%%%%%%%%%%%%%%%%%%%%
%
The homogeneity of the space in the $xy$ directions allows us to seek
the (plane wave) solutions in the form
\begin{eqnarray}
e^{-i\omega t} e^{i\vec k_\perp\cdot\vec x} \vec A(z) \,.
\end{eqnarray}
For ordinary (non-chiral) media, a suitable ansatz is
\begin{eqnarray}
\label{ansatz}
\vec A(z) = 
\left[\vec a e^{ik_\parallel z} + 
\vec b e^{-ik_\parallel z}\right]
\Theta(-z) + \left[\vec c e^{ik'_\parallel z}\right]\Theta(z) \,.
\end{eqnarray}
The form chosen in Eq.\ (\ref{ansatz}) corresponds to the physical
situation in which the wave is incident from the left (with no
reflection from the far right-hand side, as it should be in a
semi-infinite medium), as depicted in Fig.\ \ref{fig:waves}.

But here we encounter the first departure from the standard treatment 
when we consider chiral media. Namely, while we can set the
\emph{incident} component to consist of only one of the two polarizations,
the \emph{reflected} and the \emph{refracted} ones will in general
consist of a superposition of the two propagating modes, which in the
present case have different wavelengths. For example, if we decompose
the refracted wave vector into its perpendicular and parallel
components (relative to the $z = 0$ plane) $\vec k' = (\vec
k_\perp,k'_\parallel)$, then for fixed values of $\omega$ and $\vec
k_\perp$ there is actually a different value of the parallel component
for each polarization, given by $\sqrt{K^{\prime\,2}_\tau -
k^2_\perp}$, so that the two polarization components travel in
different directions.
 
With this in mind, we therefore write our proposed solution
corresponding to a wave with definite polarization $\sigma = \pm$
incident from the left, in the form
\begin{eqnarray}
\label{ansatz2}
\vec A(\vec x,t) = e^{-i\omega t} 
\left[\vec A_{I}\Theta(-z) + \vec A_{II}\Theta(z)\right] \,,
\end{eqnarray}
where $\vec A_{I}$, which contains the incident and the reflected 
components, is taken to be
\begin{eqnarray}
\label{AI}
\vec A_I & = &
a_\sigma \hat e_\sigma e^{i\vec k_\sigma\cdot\vec x } + 
\sum_{\tau = \pm} a''_\tau\hat e''_\tau 
e^{i\vec k^{''}_\tau\cdot\vec x} \,,
\end{eqnarray}
while the refracted component $\vec A_{II}$ is
\begin{eqnarray}
\label{AII}
\vec A_{II} & = & 
\sum_{\tau = \pm}a'_\tau \hat e'_\tau
e^{i\vec k'_\tau\cdot\vec x } \,.
\end{eqnarray}
The wave vectors in the expressions for these waves can be written as
\begin{eqnarray}
\label{momdecomp}
\vec k_\sigma & = & (\vec k_\perp, k_{\sigma\parallel})\,,
\nonumber\\
\vec k'_\tau & = & (\vec k_\perp, k_{\tau\parallel}')\,,
\nonumber\\
\vec k''_\tau & = & (\vec k_\perp,-k_{\tau\parallel})
\end{eqnarray}
sharing a common value of the perpendicular component.  For the same
fixed value of $\omega$, we denote by $K_\sigma(\omega)$ and
$K'_\sigma(\omega)$ the solutions of the dispersion relations in the
two regions characterized by refractive index functions $n_\pm$ and
$n'_\pm$.  Making reference to the angles shown in Fig.\
\ref{fig:waves}, the parallel components for the different wave
vectors can be written as
\begin{eqnarray}
\label{defangles}
k_{\sigma\parallel} & = & K_\sigma \cos\alpha \nonumber\\
k'_{\tau\parallel} & = & 
K'_\tau \cos\alpha'_\tau \,,\nonumber\\
k''_{\tau\parallel} & = & 
K_\tau \cos\alpha''_\tau \,,
\end{eqnarray}
where
\begin{eqnarray}
k'_{\tau\parallel} & = & \sqrt{K_\tau'{}^2 - k_\perp^2}
\,,\nonumber\\
k''_{\tau\parallel} & = & 
\sqrt{K_\tau^2 - k_\perp^2} \,.
\end{eqnarray}

In order to specify the polarization vectors uniquely, we define $\hat
e_1$ to be in the direction of $\vec k_\perp\times \hat u_z$.  Making
reference to Fig.\ \ref{fig:waves}, it points perpendicular to the
page, away from the viewer. We take that to be also the direction of
the $x$ axis, and define the $y$ axis by 
\begin{eqnarray}
\hat u_y = \hat u_z\times \hat u_x \,,
\end{eqnarray}
which is the vertical direction in the representation of Fig.\
\ref{fig:waves}.  Then, for the polarization vectors,  we take 
\begin{eqnarray}
\hat e'_1 = \hat e''_1 = \hat e_1 = \hat u_x\,,
\end{eqnarray}
while
\begin{eqnarray}
\hat e_2 & = & \hat k\times \hat e_1 \,, \nonumber\\
\hat e'_{2\tau} & = & \hat k'_\tau\times \hat e_1 \,,\nonumber\\
\hat e''_{2\tau} & = & \hat k''_\tau\times \hat e_1 \,,
\label{e2def}
\end{eqnarray}
where $\hat k$, $\hat k'_\tau$ and $\hat k''_\tau$ are
the unit vectors along $\vec k_\sigma$, $\vec k'_\tau$ and $\vec k''_\tau$,
respectively. Notice that, while the
direction of the incident wave is fixed (by us), the reflected and
refracted components consist each of two polarized waves that travel
in different directions, as we have already remarked. With the
choice of axis specified above,
\begin{eqnarray}
\hat k & = & \cos\alpha \, \hat u_z + \sin\alpha \, \hat u_y \,,\nonumber\\
\hat k'_\tau & = & \cos\alpha'_\tau \, \hat u_z + 
\sin\alpha'_\tau \, \hat u_y \,,\nonumber\\
\hat k''_\tau & = & -\cos\alpha''_\tau \, \hat u_z + 
\sin\alpha''_\tau \, \hat u_y \,,
\end{eqnarray}
and whence, using Eq.\ (\ref{e2def}),
\begin{eqnarray}
\hat e_2 & = & -\sin\alpha \, \hat u_z + \cos\alpha \, \hat u_y\,,\nonumber\\
\hat e'_{2\tau} & = &  -\sin\alpha'_\tau \, \hat u_z + 
\cos\alpha'_\tau \, \hat u_y
\,,\nonumber\\
\hat e''_{2\tau} & = & - \sin\alpha''_\tau \, \hat u_z 
-\cos\alpha''_\tau \, \hat u_y \,.
\end{eqnarray}
The circular polarization vectors that enter in Eq.\ (\ref{ansatz2}) 
are defined by
\begin{eqnarray}
\hat e_\sigma & = & \frac{1}{\sqrt2}
(\hat e_1 + i\sigma\hat e_2)\,,\nonumber\\
\hat e'_\tau & = & \frac{1}{\sqrt2}
(\hat e_{1} + i\tau\hat e'_{2\tau})\,,\nonumber\\ 
\hat e''_\tau & = & \frac{1}{\sqrt2}
(\hat e_{1} + i\tau\hat e''_{2\tau}) \,,
\end{eqnarray}
and they satisfy
\begin{eqnarray}
\label{coulombgauge}
\hat e_\sigma\cdot\vec k_\sigma & = & 0\nonumber\,,\\
\hat e''_\tau\cdot\vec k''_\tau & = & 0\nonumber\,,\\
\hat e'_\tau\cdot\vec k'_\tau & = & 0\nonumber\,.
\end{eqnarray}
Together with Eq.\ (\ref{disprel}), in this way it is ensured
that the functions
\begin{eqnarray}
e^{-i\omega t}\vec A_{X} \qquad (X = I,II) \,,
\end{eqnarray} 
satisfy the Maxwell equations in each region.

The fact that the wave vectors have the same transverse component,
as indicated in Eq.\ (\ref{momdecomp}), 
implies the familiar relationship (Snell's law) between the angles of
incidence and refraction,
\begin{eqnarray}
\label{snell}
\frac{\sin\alpha'_\tau}{\sin\alpha} = \frac{K_\sigma}{K'_\tau} = 
\frac{n_\sigma}{n'_\tau} \,,
\end{eqnarray}
and the analogous relation for the reflected wave
\begin{eqnarray}
\label{reflectioncondition}
\alpha''_\tau & = & \alpha \qquad (\tau = \sigma) \nonumber\\
\frac{\sin\alpha''_\tau}{\sin\alpha} = \frac{K_\sigma}{K_\tau}
& = & \frac{n_\sigma}{n_\tau} 
\qquad (\tau \not= \sigma) \,.
\end{eqnarray}
Thus, for a given value of $\omega$ and $\alpha$, so that the wave
vector $\vec k_\sigma$ of the incident component is completely
specified, the wave vectors $\vec k''_\tau$ and $\vec k'_\tau$ of the
reflected and refracted components are also completely determined, as
well as the polarization vectors.  The only quantities yet to be
determined are the amplitudes $a'_\tau,a''_\tau$ in Eq.\
(\ref{ansatz2}).

From Eqs.\ (\ref{snell}) and (\ref{reflectioncondition})
the familiar effects such as \emph{total internal reflection}
can be deduced, in the form that they apply to the present situation.
Thus, for example,
if the dielectric constants are such that $K_\sigma > K'_\tau$,
there is a maximum incident angle $\alpha_{\tau\,{\rm max}}$,
determined by
\begin{eqnarray}
\label{totalref}
(\sin\alpha_\tau)_{\rm max} = \frac{K'_\tau}{K_\sigma} \,,
\end{eqnarray}
above which the refracted wave with polarization $\tau$ propagates parallel to
the surface ($k'_{\tau\parallel} = 0)$ but does not penetrate
the region $II$. 

In general, the picture that emerges is the following.  The incident
wave, which has a definite circular polarization, is split at the
boundary into a reflected and refracted wave, each of which is a
superposition of the two circular polarizations. However, the two
circularly polarized modes do not travel in the same direction
as a consequence of the fact that their wave vectors
have the same transverse component but a different longitudinal one.
The determination of the relative admixture of the two
polarization modes in the reflected and the refracted wave is
precisely one of the dynamical questions that we will address below.

\subsection{Dynamics}
Some of the dynamical issues that we wish to consider involve
(i) finding the amplitudes of the reflected and refracted
components of the wave for a given incident amplitude;
(ii) determining how the polarization of the wave changes as it
crosses the boundary. 
In contrast with the kinematical properties that we considered in the
previous section, these dynamical aspects depend on the specific
nature of the boundary conditions at the interface. 
The latter are determined by the requirement
that the solution given in Eq.\ (\ref{ansatz2})
satisfies the Maxwell equations at the interface.

%%%%%%%%%%%%%%%%%%%%
\subsubsection{Boundary Conditions}
%%%%%%%%%%%%%%%%%%%%
As we have already mentioned, the dispersion relations given in Eq.\
(\ref{disprel}) together with the conditions given 
in Eqs.\ (\ref{coulombgauge}) imply that the function $\vec A(\vec x,t)$
defined in Eq.\ (\ref{ansatz2}) satisfies the Maxwell equations in
each region separately. However, we have yet to ensure that the
equations are satisfied at the boundary between the two regions
itself.  This requirement yields further conditions.  Are they the
same as for the ordinary equations? Or does the $\zeta$ activity
constant term modify them in any way?  Answering this question is our
next hurdle.

To begin, we calculate the electric and magnetic field associated
with the vector potential given in Eq.\ (\ref{ansatz2}).
The standard formulas 
\begin{eqnarray}
\vec E & = & -\frac{\partial\vec A}{\partial t}\,, \nonumber\\
\vec B & = & \vec \nabla\times\vec A \,,
\end{eqnarray}
yield
\begin{eqnarray}
\label{E}
\vec E & = & e^{-i\omega t}
\left[\vec E_{I}\Theta(-z) + \vec E_{II}\Theta(z)\right] \,,\\
\label{B}
\vec B & = & e^{-i\omega t}\left[\vec B_I\Theta(-z) + \vec B_{II}\Theta(z)
+ \delta(z)\hat u_z\times(\vec A_{II} - \vec A_I)\right] \,,
\end{eqnarray}
where
\begin{eqnarray}
\vec E_X & \equiv & i\omega A_X \,,\nonumber\\
\vec B_X & \equiv & \vec\nabla\times \vec A_X \qquad (X = I,II) \,.
\label{EBX}
\end{eqnarray}
Since the fields must be finite everywhere (otherwise the Maxwell
equations would be not be satisfied), the last term in Eq.\ (\ref{B})
must be zero, which yields the condition
\begin{eqnarray}
\label{tangE}
\hat u_z\times \left[a_\sigma\hat e_\sigma + 
\sum_\tau(a''_\tau\hat e''_\tau - a'_\tau\hat e'_\tau)
\right] = 0 
\end{eqnarray}
on the polarization vectors. 

The remaining relations follow from the Maxwell equations themselves
which were given in Eqs.\ (\ref{divB}), (\ref{faraday}), (\ref{gauss})
and (\ref{ampere}) in the Introduction.  The argument to obtain the
implied conditions is similar to the one used to arrive at Eq.\
(\ref{tangE}). For example, consider Eq.\ (\ref{gauss}).  Writing
\begin{eqnarray}
\hat \epsilon \vec E = e^{-i\omega t}
\left[\epsilon \vec E_{I}\Theta(-z) + 
\epsilon' \vec E_{II}\Theta(z)\right] \,,
\end{eqnarray}
we then calculate
\begin{eqnarray}
\label{divepsilonE}
\vec\nabla\cdot \Big(\hat\epsilon \vec E \Big) 
 =  e^{-i\omega t}
\left[\Theta(-z)\vec\nabla\cdot (\epsilon\vec E_{I}) + 
\Theta(z)\vec\nabla\cdot (\epsilon' \vec E_{II})\right.
\nonumber\\*  
+ \left.\delta(z)\hat u_z\cdot
\left(\epsilon' \vec E_{II} - \epsilon \vec
E_I\right) \right] \,.
\end{eqnarray}
The first two terms on the right-hand side are identically zero 
since there is no free charge in either region. Thus, the condition
that the divergence of $\hat\epsilon \vec E$ is zero everywhere
requires that the coefficient of $\delta(z)$ in Eq.\ (\ref{divepsilonE}) 
be zero. This yields the relation
\begin{eqnarray}
\label{perpD}
\hat u_z\cdot\left[\epsilon a_\sigma\hat e_\sigma + 
\sum_\tau\left(\epsilon a''_\tau\hat e''_\tau - 
\epsilon' a'_\tau\hat e'_\tau\right)\right] = 0 \,,
\end{eqnarray}
where we have used the expressions for $\vec E_{I},\vec E_{II}$ that
follow from Eqs.\ (\ref{ansatz2}), (\ref{AI}), (\ref{AII}) and
(\ref{EBX}).

The conditions that follow from the remaining Maxwell equations are
obtained similarly. In general, by means of Eqs.\ (\ref{disprel}) and
the definition of the corresponding polarization vectors given above,
it is ensured that the proposed solution satisfies
the equations in each region separately. Thus, when the solution is
substituted in the Maxwell equations, only the terms that arise from
the derivatives acting on the Theta function, which are proportional
to the delta function, are not automatically zero.  Demanding that
they be zero as well yields the required relations.  Proceeding
in this way we find that Eq.\ (\ref{faraday}) yields Eq.\ (\ref{tangE}) again, 
and the remaining two conditions that
complement Eqs. (\ref{tangE}) and (\ref{perpD}) are
\begin{eqnarray}
\label{perpB}
\hat u_z\cdot\left[\sigma a_\sigma \hat e_\sigma K_\sigma + \sum_\tau\tau
\left(K_\tau a''_\tau\hat e''_\tau - K'_\tau
a'_\tau\hat e'_\tau\right)\right] & = & 0 \,,\\
\label{tangH}
\hat u_z\times\left[\sigma a_\sigma \epsilon v_\sigma 
\hat e_\sigma 
+ \sum_\tau \tau \left( a''_\tau \epsilon v_\tau 
\hat e''_\tau
- a'_\tau \epsilon' v'_\tau
\hat e'_\tau \right) \right] & = & 0 \,.
\end{eqnarray}
In writing these conditions, we have used the relation
\begin{eqnarray}
\label{Sk}
i\hat k_\sigma \times \hat e_\sigma = \sigma \hat e_\sigma
\end{eqnarray}
and the definition of $v_\sigma$ from Eq.\ (\ref{npm}), 
as well as the analogous ones for the corresponding primed and doubly-primed
quantities.
In addition we have used the fact that, for any wave
(incident, refracted and reflected), the product of
its wave number times its velocity
velocity equals $\omega$, which is the same for all of them.

%
% section 5
%
\section{Solution to the boundary conditions}
\label{sec:solutions}
\subsection{The independent conditions}
Our task is to use the conditions given in Eqs.\ (\ref{tangE}),
(\ref{perpD}), (\ref{perpB}) and (\ref{tangH}) to obtain, for a given
incident amplitude $a_\sigma$, the reflected and transmitted
amplitudes $a'_\tau,a''_\tau$.  There are thus a total of four unknown
variables to be solved for.  On the other hand, at first sight, the
number of conditions seems to be six.  While Eqs.\ (\ref{perpD}) and
(\ref{perpB}) are scalar equations, the relations in Eqs.\
(\ref{tangE}) and (\ref{tangH}) are vectorial.  Being a
two-dimensional problem, as we have seen, each of the vector relations
yields two conditions, for a total of six. Therefore, the solution for
$a'_\tau,a''_\tau$ is algebraically overdetermined, and a non-trivial
solution exists only if additional auxiliary conditions are
satisfied. As we will see, these are just the Snell-like relations
given in Eqs.\ (\ref{snell}) and (\ref{reflectioncondition}), and the
dispersion relations of Eq.\ (\ref{disprel}). Put in another way, when
Eq.\ (\ref{disprel}) as well as Eqs.\ (\ref{snell}) and
(\ref{reflectioncondition}) are satisfied, two of the six relations
implied by the set of boundary conditions are redundant, leaving just
four independent equations for the four unknowns $a'_\tau,a''_\tau$.

Before going to the details of the problem, it is useful to recollect
the corresponding situation in the case of ordinary, non-chiral media.
In that case, for a given value of the transverse component of the
wave vector and a given frequency, the longitudinal component of the
wave vector in each of the reflected and refracted waves has a unique
value.  Thus, the solution written down in Eq.\ (\ref{ansatz2}) indeed
collapses to the form given in Eq.\ (\ref{ansatz}). Nevertheless, the
problem is similar to the one stated above.  Namely, for a given
incident amplitude $\vec a$, we wish to know what are the the
coefficients $\vec b$ and $\vec c$.  Being a two-dimensional problem
as already argued, then for an arbitrary (but definite) $\vec a$ in
the $\hat e_{1,2}$ plane, there are four unknowns, represented by the
two components of $\vec b$ in the $\hat e''_{1,2}$ plane and the two
components of $\vec c$ in the $\hat e'_{1,2}$ plane.  Further, the
problem being a linear one (the boundary conditions are linear
equations in the amplitudes), we can proceed to find the particular
solutions corresponding to the case $\vec a = \hat e_1$ and,
separately, $\vec a = \hat e_2$, and the general solution for an
arbitrary choice of $\vec a$ is obtained by linear superposition of
those particular ones. This is actually the procedure followed in
Jackson's book \cite{jackson}.  The fundamental reason why a solution
can be obtained at all, is the fact that there are four independent
algebraic relations (that follow from the boundary conditions) among
the four unknowns the we have identified.

Returning to our problem, these same general principles hold as well.
Namely, while some features and details of the solutions are not
applicable or relevant, the linearity property, superposition
principle and the fact that the problem is algebraically well defined,
also apply.  The fundamental difference, as far as the algebraic
details and manipulations is concerned, is that in the present case it
is not useful to build up the general solution by superposing the
particular solutions obtained for the cases $\vec a = \hat e_{1,2}$,
and decomposing the reflected and refracted waves in terms of the
linear polarization components. In our case, the circular polarization
basis is, as we have shown, the appropriate one to use.

For the remaining vector algebra we use the relations
\begin{eqnarray}
\hat u_z \cdot \hat e_\sigma & = & 
-\frac{1}{\sqrt{2}}i\sigma\sin\alpha\,,\nonumber\\
\hat u_z \cdot \hat e^\prime_\tau & = & 
-\frac{1}{\sqrt{2}}i\tau\sin\alpha^\prime_\tau\,,\nonumber\\
\hat u_z \cdot \hat e''_\tau & = & 
-\frac{1}{\sqrt{2}}i\tau\sin\alpha''_\tau\,,
\end{eqnarray}
and
\begin{eqnarray}
\hat u_z\times\hat e_\sigma & = & \frac{1}{\sqrt{2}}\left(
\hat u_y - i\sigma \cos\alpha\hat u_x\right)\,,\nonumber\\
\hat u_z\times\hat e^\prime_\tau & = & \frac{1}{\sqrt{2}}\left(
\hat u_y - i\tau \cos\alpha^\prime_\tau\hat u_x\right)\,,\nonumber\\
\hat u_z\times\hat e''_\tau & = & \frac{1}{\sqrt{2}}\left(
\hat u_y + i\tau \cos\alpha''_\tau\hat u_x\right)\,.
\end{eqnarray}
Using these multiplication rules, Eqs.\ (\ref{tangE}), (\ref{perpD}),
(\ref{perpB}) and (\ref{tangH}) imply, in that order, the following
six relations
\begin{eqnarray}
\label{thesixequations}
\sum_\tau(a'_\tau - a''_\tau) & = & a_\sigma \,,\nonumber\\
\sum_\tau \tau
(a'_\tau\cos\alpha'_\tau + 
a''_\tau\cos\alpha''_\tau) & = & \sigma a_\sigma\cos\alpha
\,,\nonumber\\ 
\sum_\tau (K'_\tau a'_\tau\sin\alpha'_\tau - K_\tau
a''_\tau\sin\alpha''_\tau) & = & a_\sigma 
K_\sigma\sin\alpha \,,\nonumber\\ 
\sum_\tau \tau(\epsilon' a'_\tau\sin\alpha'_\tau - \epsilon
a''_\tau\sin\alpha''_\tau) & = & \sigma a_\sigma\epsilon \sin\alpha
 \,,\nonumber\\
\sum_\tau \tau\left(
a'_\tau \epsilon' v'_\tau -
a''_\tau \epsilon v_\tau
\right) & = & \sigma
a_\sigma \epsilon v_\sigma \,,\nonumber\\ 
\sum_\tau \left(
a'_\tau \epsilon' v'_\tau \cos\alpha'_\tau +
a''_\tau \epsilon v_\tau \cos\alpha''_\tau 
\right) & = & a_\sigma \epsilon v_\sigma \cos\alpha \,.
\end{eqnarray}
If we now use Snell's law [Eqs.\ (\ref{snell}) and
(\ref{reflectioncondition})], it is easy to see that the third
equation becomes identical to the first, and the fourth identical to
the fifth.  Thus, we are left with four independent equations which,
for easy reference in what follows, we recollect below:
\begin{eqnarray}
\label{indepeqns}
\sum_\tau(a'_\tau - a''_\tau) & = & a_\sigma \,,\nonumber\\*
\sum_\tau \tau
(a'_\tau\cos\alpha'_\tau + 
a''_\tau\cos\alpha''_\tau) & = & \sigma a_\sigma\cos\alpha
\,,\nonumber\\*
\sum_\tau \tau\left(
a'_\tau \epsilon' v_\tau' -
a''_\tau \epsilon v_\tau
\right) & = & \sigma
a_\sigma \epsilon v_\sigma \,,\nonumber\\ 
\sum_\tau \left(
a'_\tau \epsilon' v_\tau' \cos\alpha'_\tau +
a''_\tau \epsilon v_\tau \cos\alpha''_\tau 
\right) & = & a_\sigma \epsilon v_\sigma \cos\alpha \,.
\end{eqnarray}
These equations must be solved for the four unknowns $a'_\tau$ and
$a''_\tau$ (with $\tau=\pm$), for any given incident amplitude
$a_\sigma$. Before embarking on the general solution, it is
instructive to consider various particular cases.

%%%%%%%%%%%%%%%%%%%%
\subsection{Non-chiral media as a special limiting case}
\label{subsec:ordinarymedia}
%%%%%%%%%%%%%%%%%%%%
If $\zeta$ is zero in both sides of the boundary, so that we are
referring to ordinary, non-chiral, media, the equations for
the coefficients simplify. Neither the indices of refraction,
nor the angles, depend on the sign of the polarization.  In terms of
the linear polarization amplitudes
\begin{eqnarray}
a_1 & = & \frac{1}{\sqrt{2}}(a_{+} + a_{-})\,,\nonumber\\
a_2 & = & \frac{i}{\sqrt{2}}(a_{+} - a_{-})\,,
\end{eqnarray}
and similar ones for the primed and doubly primed coefficients,
the equations become
\begin{eqnarray}
\label{nonchiralconditions}
a'_1 - a''_1 & = & \frac{1}{\sqrt{2}}a_\sigma \,,\nonumber\\
a'_2\cos\alpha' + a''_2\cos\alpha & = & 
\frac{i}{\sqrt{2}}\sigma a_\sigma\cos\alpha \,,\nonumber\\
a'_2\epsilon' v' - a''_2\epsilon v & = & 
\frac{i}{\sqrt{2}}\sigma a_\sigma \epsilon v\,,\nonumber\\
a'_1\epsilon' v'\cos\alpha' +
a''_1\epsilon v\cos\alpha & = & 
\frac{1}{\sqrt{2}}\sigma a_\sigma \epsilon v\cos\alpha\,.
\end{eqnarray}
Thus, the set of four equations splits into two $2\times 2$ blocks,
one for $a'_1$ and $a''_1$ and the other for $a'_2$ and $a''_2$, which
are easily solved to give
\begin{eqnarray}
\label{nonchiralsolution12}
\left(
\begin{array}{c}
a'_1 \\ a''_1
\end{array}\right) & = & \frac{1}{\sqrt{2}}a_\sigma E_1 \nonumber\\[12pt]
\left(
\begin{array}{c}
a'_2 \\ a''_2
\end{array}\right) & = & \frac{i}{\sqrt{2}}\sigma a_\sigma E_2 \,,
\end{eqnarray}
where
\begin{eqnarray}
\label{nonchiralsolution12E}
E_1 & = &
\frac{1}{\epsilon v\cos\alpha + \epsilon' v'\cos\alpha'}
\left(
\begin{array}{c}
2\epsilon v\cos\alpha \\[12pt] 
\epsilon v\cos\alpha - \epsilon' v'\cos\alpha'
\end{array}\right) \nonumber\\[12pt]
E_2 & = &
\frac{1}{\epsilon v\cos\alpha' + \epsilon' v'\cos\alpha}
\left(
\begin{array}{c}
2\epsilon v\cos\alpha \\[12pt] 
\epsilon' v'\cos\alpha - \epsilon v\cos\alpha'
\end{array}\right) \,.
\end{eqnarray}
The original coefficients for the circularly polarized amplitudes,
\begin{eqnarray}
a'_\tau = \frac{1}{\sqrt{2}}(a'_1 - i\tau a'_2)
\end{eqnarray}
and similarly for $a''_\tau$, are then given by
\begin{eqnarray}
\label{nonchiralsolutionpm}
\left(
\begin{array}{c}
a'_\tau \\ a''_\tau
\end{array}\right) = \frac{1}{2}a_\sigma(E_1 + \sigma\tau E_2) \,.
\end{eqnarray}

Regarding the physical interpretation, the formulas in Eq.\
(\ref{nonchiralsolution12}) give the solution for the reflected and
refracted amplitudes expressed in the linear polarization basis, with
the corresponding quantities in the circular polarization basis being
obtained by means of Eq.\ (\ref{nonchiralsolutionpm}). It should be
noted further that, whichever basis we use to express the solution, we
have assumed that the incident wave has a definite circular
polarization specified by $\sigma$, which can be $\pm 1$.  However, we
can easily obtain the solutions for an arbitrary polarization of the
incident wave by superposition.  If the incident wave is a combination
of the two circularly polarized waves, then the solution is
generalized to
\begin{eqnarray}
\label{nonchiralsolutionpmgen}
\left(
\begin{array}{c}
a'_\tau \\ a''_\tau
\end{array}\right) = \frac{1}{2}\sum_\sigma a_\sigma(E_1 + \sigma\tau E_2) \,.
\end{eqnarray}
or equivalently, in the linear basis,
\begin{eqnarray}
\label{nonchiralsolution12gen}
\left(
\begin{array}{c}
a'_1 \\ a''_1
\end{array}\right) & = & \frac{1}{\sqrt{2}}\sum_\sigma
a_\sigma E_1 = a_1 E_1\nonumber\\[12pt]
\left(
\begin{array}{c}
a'_2 \\ a''_2
\end{array}\right) & = & \frac{i}{\sqrt{2}}\sum_\sigma \sigma a_\sigma E_2 =
a_2 E_2\,.
\end{eqnarray}

This solution in fact reproduces the well known results for the situation
we are considering. It is reassuring to confirm, for example, that
it embodies the Fresnel formulas
when we take the incident wave to be linearly polarized
along $\hat u_x$ or $\hat u_y$.
Let us consider the case in which the incident wave
is linearly polarized along the $\hat u_x$ direction, which implies that
the incident wave is an equal admixture of the two circularly
polarized waves, with the coefficients satisfying
\begin{eqnarray}
\label{linearpolincident}
a_{+} = a_{-} & \equiv & \frac{1}{\sqrt{2}}a_1 \,,\nonumber\\
a_2 & = & 0\,.
\end{eqnarray}
Then the solution for this case, namely
\begin{eqnarray}
\label{soltransverseE}
\left(
\begin{array}{c}
a'_1 \\ a''_1
\end{array}\right) & = & a_1 E_1 \,,\nonumber\\[12pt]
\left(
\begin{array}{c}
a'_2 \\ a''_2
\end{array}\right) & = & 0 
\end{eqnarray}
in the linear basis, is immediately
recognized as the so-called \emph{Transverse Electric\/} solution.
Similarly, the solution for
linear polarization along $\hat e_2$,
\begin{eqnarray}
\label{soltransverseB}
\left(
\begin{array}{c}
a'_1 \\ a''_1
\end{array}\right) & = & 0 \,,\nonumber\\[12pt]
\left(
\begin{array}{c}
a'_2 \\ a''_2
\end{array}\right) & = & a_2 E_2 
\end{eqnarray}
corresponds to the standard \emph{Transverse Magnetic} solution.
The Fresnel formulas are reproduced by writing the above
results in terms of the magnetic permeability $\mu$ and the refractive
index $n$, instead of quantities $\epsilon$ and $v$.
These two sets of quantities are related by Eq.\ (\ref{npm}),
which for non-chiral media reads
\begin{eqnarray}
v^2 \equiv {1 \over n^2} = \frac{1}{\epsilon\mu} \,.
%\label{}
\end{eqnarray}
Eliminating $\epsilon$ and $v$ in favor of $\mu$ and $n$, 
the transverse electric solution can be written as
\begin{eqnarray}
{a'_1 \over a_1} &=& {2n \cos\alpha \over n \cos\alpha + {\mu\over
\mu'} n' \cos\alpha'}\,, \nonumber\\*
{a''_1 \over a_1} &=& {n \cos\alpha - {\mu\over
\mu'} n' \cos\alpha' \over n \cos\alpha + {\mu\over
\mu'} n' \cos\alpha'}\,,
%\label{}
\end{eqnarray}
which, after using Snell's law to eliminate the angle of refraction,
is the usual form (given, for example, by Jackson\cite{jackson}).
The result for the transverse magnetic case can be reproduced similarly.

%%%%%%%%%%%%%%%%%%%%
\subsection{Incidence from ordinary to chiral media}
%%%%%%%%%%%%%%%%%%%%
A particularly simple situation, which reveals some of the features of
the general solution, arises by considering the case in which the
first medium is an ordinary dielectric, but the second one is not, i.e., 
\begin{eqnarray}
\zeta = 0\,, \qquad \zeta' \not= 0 \,.
\end{eqnarray}
In this case,
\begin{eqnarray}
n_{+} = n_{-} \equiv n
\end{eqnarray}
and by Snell's law
\begin{eqnarray}
\alpha''_{+} = \alpha''_{-} = \alpha\,,
\end{eqnarray}
so that the equations for the coefficients become
\begin{eqnarray}
(a'_+ + a'_-) - (a''_+ + a''_-) & = & a_\sigma \,,\nonumber\\
(a'_+\cos\alpha'_+ - a'_-\cos\alpha'_-) + 
(a''_+ - a''_-)\cos\alpha & = & \sigma a_\sigma\cos\alpha \,,\nonumber\\
\epsilon'(a'_+v'_+ - a'_-v'_-) - (a''_+ - a''_-)\epsilon v
& = & \sigma a_\sigma\epsilon v \,,\nonumber\\
\epsilon' (a'_+v'_+\cos\alpha'_{+} + a'_-v'_-\cos\alpha'_{-}) 
+ (a''_+ + a''_-)\epsilon v\cos\alpha
& = & a_\sigma\epsilon v\cos\alpha \,.
\end{eqnarray}
We can eliminate the doubly-primed coefficients in the fourth using
the first, and likewise in the third using the second. The resulting
equations are easily solved for the primed coefficients, and the
doubly primed ones can be obtained by using the first and the second
equations. Thus, the solution is easily obtained, and it can be
expressed in the form
\begin{eqnarray}
\frac{a'_\tau}{a_\sigma} & = & 
\left(\frac{2\epsilon v\cos\alpha}{T}\right)
T^\sigma_\tau\,,\nonumber\\*[12pt] 
\frac{a''_\tau}{a_\sigma} & = & 
-\left(\frac{1 - \tau\sigma}{2}\right) +
\frac{\epsilon v}{T}\sum_{\lambda = \pm}
\left(\cos\alpha - \lambda\tau\cos\alpha'_\lambda\right)
T^\sigma_\lambda \,, 
\end{eqnarray}
where
\begin{eqnarray}
\label{defTsigmatau}
T^\sigma_\tau = 
\left(\epsilon v + \tau\sigma\epsilon' v'_{-\tau}\right)
\left(\cos\alpha'_{-\tau} + \tau\sigma\cos\alpha\right) \,,
\end{eqnarray}
and
\begin{eqnarray}
T & \equiv & \frac{1}{2}\left(
T^{+}_{+} T^{-}_{-} - T^{+}_{-} T^{-}_{+}\right) \nonumber\\* 
& = & \epsilon\epsilon' v(v'_{+} + v'_{-})
(\cos^2\alpha + \cos\alpha'_{+}\cos\alpha'_{-}) \nonumber\\* 
&& +
\left(\epsilon^2 v^2 + \epsilon^{\prime\,2}v'_{+}v'_{-}\right)
\cos\alpha(\cos\alpha'_{+} + \cos\alpha'_{-}) \,.
\end{eqnarray}
%

%%%%%%%%%%%%%%%%%%%%
\subsubsection*{Normal incidence}
%%%%%%%%%%%%%%%%%%%%
A particularly simple solution is obtained if,
in addition, we assume that the incident wave travels normal to the
plane boundary, $\alpha = 0$, in which case the other angles are also
all equal to $0$ by Snell's law. The above solution then reduces to
\begin{eqnarray}
\frac{a'_\tau}{a_\sigma} & = & 
\frac{1}{2}(1 + \tau\sigma) F_\tau \,,\nonumber\\
\frac{a''_\tau}{a_\sigma} & = & 
\frac{1}{2}(1 - \tau\sigma) G_{-\tau} \,,
\end{eqnarray}
where we have defined
\begin{eqnarray}
F_\tau & = & \frac{2\epsilon v}{\epsilon v + \epsilon' v'_\tau}\,,
\nonumber\\[12pt]
G_\tau & = & F_\tau - 1 = \frac{\epsilon v - \epsilon' v'_\tau}
{\epsilon v + \epsilon' v'_\tau}\,.
\end{eqnarray}

This solution shows various interesting properties. For example, if we
consider either $\sigma = +$ or $\sigma = -$, the transmitted wave has
the same circular polarization as the incident one, but the reflected
wave has the opposite. This is a well known effect even in the case of
ordinary media ($\zeta' = 0$). The distinguishing feature in the
present case is that the intensity of the
reflected (and the transmitted) wave
depends on what is the polarization of the incident
one.

If the incident wave does not have a definite circular polarization,
but it is a linear combination of them, then by superposition
the general solution for that case is
\begin{eqnarray}
\frac{a'_\tau}{a_\sigma} & = & \frac{1}{2}F_\tau
\sum_\sigma(1 + \tau\sigma) \,,\nonumber\\
\frac{a''_\tau}{a_\sigma} & = & 
\frac{1}{2}G_{-\tau}\sum_\sigma(1 - \tau\sigma)  \,.
\end{eqnarray}
Let us consider the case of linear polarization along $\hat e_1$,
as in Eq.\ (\ref{linearpolincident}). In that case the solution is
\begin{eqnarray}
\frac{a'_\tau}{a_1} & = & \frac{1}{\sqrt{2}} F_\tau\,,\nonumber\\
\frac{a''_\tau}{a_1} & = & \frac{1}{\sqrt{2}} G_{-\tau}\,.
\end{eqnarray}
Thus neither the reflected nor the transmitted waves have a definite linear
polarization (e.g., $a'_+\not= a'_-$)
in contrast to the situation considered in 
Section\ \ref{subsec:ordinarymedia}.
Furthermore, while the incident wave is represented by
\begin{eqnarray}
\vec A^{(\rm inc)}_I(z) = a_1\hat e_1 e^{iKz} \,,
\end{eqnarray}
the transmitted component is
\begin{eqnarray}
\vec A_{II}(z) = \frac{1}{\sqrt{2}}a_1\left[
\hat e'_+F_+e^{iK'_+z}
+ \hat e'_-F_-e^{iK'_-z}\right]\,.
\end{eqnarray}
That is, while the incident wave is, at the boundary ($z = 0$), an
equal admixture of the positive and negative circular polarizations,
given by the $\hat e_1$ direction, the transmitted wave is split into
an unequal admixture of the two circular polarizations.  
As we saw in Section\ \ref{subsec:ellipticalwaves}, such an admixture
describes an elliptically polarized wave, with the particular
characteristics that are due to the chiral nature of the medium,
as we explained there. Moreover,
in the region $z > 0$, the vector that determines the
direction of the electric field in the present case
is obtained from the formula given in Eq.\ (\ref{poldirectionellip}),
by making the substitutions
\begin{eqnarray}
A_{\pm} & \rightarrow & F_{\pm} \,,\nonumber\\
\theta(z) & \rightarrow & 
\frac{1}{2}(K^\prime_{-} - K^\prime_{+})z \,,\nonumber\\
\Delta(z) & \rightarrow & \frac{1}{2}(K^\prime_{+} + K^\prime_{-})z \,.
\end{eqnarray}
%

%%%%%%%%%%%%%%%
\subsection{General solution}
%%%%%%%%%%%%%%%
Other special cases can be treated similarly, such as the case of
incidence from chiral to ordinary media, including the particular
situation of normal incidence.  However, we do not proceed any further
along those lines, and go directly to give the solution to
Eq.\ (\ref{indepeqns}) for the reflected and refracted
amplitudes, in the general case.  
For this purpose, we use the shorthand notation
$c_0 \equiv \cos\alpha$, $c'_{\pm}\equiv\cos\alpha'_{\pm}$ and
$c^{\prime\prime}_{\pm} \equiv \cos\alpha^{\prime\prime}_{\pm}$ for
the cosine of various angles. Then, defining
\begin{eqnarray}
D &=& \phantom+
      (c'_+ c'_- + c''_+ c''_-) 
(\epsilon' v'_+ + \epsilon' v'_-) 
(\epsilon v_+ + \epsilon v_-)
\nonumber\\* && 
+     (c'_+ c''_- + c'_- c''_+) 
(\epsilon' v'_- + \epsilon v_+) 
(\epsilon' v'_+ + \epsilon v_-) 
\nonumber\\* && 
+     (c'_+ c''_+ + c'_- c''_-) 
(\epsilon' v'_+ - \epsilon v_+) 
(\epsilon' v'_- - \epsilon v_-) \,,
%\label{}
\end{eqnarray}
the solution is given by
\begin{eqnarray}
{a'_\tau \over a_\sigma} \, D
&=& 
\phantom+ (\sigma \epsilon v_\sigma + \tau \epsilon' v'_{-\tau}) 
(\epsilon v_+ + \epsilon v_-) 
(\sigma c_0 c'_{-\tau} + \tau c''_+ c''_-)
\nonumber\\* &&
+ (\sigma \epsilon v_\sigma + \epsilon v_-) 
(\epsilon' v'_{-\tau} + \tau \epsilon v_+) 
(\sigma c_0 c''_- + \tau c'_{-\tau} c''_+)
\nonumber\\* &&
+ (\sigma \epsilon v_\sigma - \epsilon v_+) 
(\epsilon' v'_{-\tau} - \tau \epsilon v_-) 
(\sigma c_0 c''_+ + \tau c'_{-\tau} c''_-)
\,, \\
%%%
{a''_\tau \over a_\sigma}  \, D
&=& 
\phantom+ (\sigma \epsilon v_\sigma + \tau \epsilon v_{-\tau}) 
(\epsilon' v'_+ + \epsilon' v'_-) 
(\sigma c_0 c''_{-\tau} - \tau c'_+ c'_-) 
\nonumber\\* &&
+ (\sigma \epsilon v_\sigma + \tau \epsilon' v'_{-\tau}) 
(\epsilon' v'_\tau + \epsilon v_{-\tau}) 
(\sigma c_0 c'_{-\tau} - \tau c'_\tau c''_{-\tau})
\nonumber\\* &&
- (\sigma \epsilon v_\sigma - \tau \epsilon' v'_\tau) 
(\epsilon' v'_{-\tau} - \epsilon v_{-\tau}) 
(\sigma c_0 c'_\tau - \tau c'_{-\tau} c''_{-\tau}) \,.
%\label{}
\end{eqnarray}
These formulas can be written in several alternate forms,
in terms of the angle of incidence and the refractive indices, by
using Snell's law [Eq.\ (\ref{snell})]. In the appropriate limits,
they reduce to the special case formulas already seen.

%
% section 6
%
\section{Conclusions}
\label{sec:conc}
The electromagnetic properties of a medium that exhibits chirality
(also called optical activity), but which is otherwise linear,
homogeneous and isotropic, can be described in terms of the usual two
parameters $\epsilon$ and $\mu$ that represent the dielectric and
permeability functions, and an additional parameter $\zeta$ that is
indicative of the chirality property.  As we have argued here and in
the references cited, such a parameterization in terms of just one
additional function, is both complete and minimal.  In the case of an
infinite (unbounded) medium, an electromagnetic wave exhibits the
phenomenon known as `natural optical activity', which is due to the
fact that the two circularly polarized states travel with different
speed if $\zeta$ is non-zero.

In this work we have considered in detail various aspects of the
propagation of a wave, in a medium that is made of two semi-infinite
media, either or both of which may be chiral, which are separated by a
plane interface.  Mimicking the procedure that is applied to the
analogous problem involving ordinary dielectrics, our approach was
based on writing the plane wave solution that satisfies the Maxwell
equations in each region, supplemented by the appropriate boundary
conditions at the interface.

We considered first various kinematical aspects of the solution which
are independent of the detailed nature of the boundary conditions. In
particular, we obtained the Snell law and the condition for the total
internal reflection effect, in the form that it applies to the present
situation.

In contrast with the above, the determination of the relative
amplitudes of the reflected and refracted components is a dynamical
issue that involve the details of the boundary conditions.  A
significant problem that we had to solve was precisely to elucidate
what is the appropriate set of boundary conditions that must be
satisfied at the interface.  The boundary conditions depend on the
Maxwell equations themselves, and since these depend on the parameter
$\zeta$, the boundary conditions necessarily involve that parameter as
well.

After finding the appropriate set of independent conditions, we solved
them and obtained the amplitudes in various particular cases and
configurations, such as one region being chiral and the other one
being an ordinary dielectric, and the case of normal incidence.  We
considered various features of those solutions, and in particular we
obtained in various cases the corresponding formula for the angle of
the direction of polarization.

In the general case, the boundary conditions form a set of four
(linear) equations that must be solved for the four unknown
amplitudes in terms of the amplitude of the incident wave.
While the explicit form of the solution is not particularly
illuminating, we obtained it and for completeness we wrote it down.

Our work opens the way for handling a whole class of related problems
involving chiral media, that we can now formulate in a concrete way.
We can consider, for example, a slab of one material of finite
thickness, inserted between two semi-infinite media, and in various
configurations depending on which ones are chiral or ordinary
dielectrics.  Possible generalizations include also periodic or
semi-periodic arrays of alternating materials, and similar
arrangements.  {From} an algebraic point of view, all of them will be
ultimately reduced to writing down the plane wave solutions in each
region, and imposing the boundary conditions at each interface.  In
this work we have shown the way for treating and solving all such
problems systematically.  In addition to the interesting and
potentially important physical applications that we have mentioned,
some of the ideas exposed here can be useful in other physics
problems that have been studied in the literature
\cite{Hehl:2002hr,Balakin:2002mc}, which have a similar mathematical
structure.

\begin{acknowledgments}
This work was partially supported (JFN) by the U.S. National
Science Foundation Grant PHY-0139538.
\end{acknowledgments}


\begin{thebibliography}{19}
\expandafter\ifx\csname natexlab\endcsname\relax\def\natexlab#1{#1}\fi
\expandafter\ifx\csname bibnamefont\endcsname\relax
  \def\bibnamefont#1{#1}\fi
\expandafter\ifx\csname bibfnamefont\endcsname\relax
  \def\bibfnamefont#1{#1}\fi
\expandafter\ifx\csname citenamefont\endcsname\relax
  \def\citenamefont#1{#1}\fi
\expandafter\ifx\csname url\endcsname\relax
  \def\url#1{\texttt{#1}}\fi
\expandafter\ifx\csname urlprefix\endcsname\relax\def\urlprefix{URL }\fi
\providecommand{\bibinfo}[2]{#2}
\providecommand{\eprint}[2][]{\url{#2}}

\bibitem[{\citenamefont{Jackson}(1975)}]{jackson}
\bibinfo{author}{\bibfnamefont{J.~D.} \bibnamefont{Jackson}},
  \emph{\bibinfo{title}{Classical Electrodynamics, 2nd edition}}
  (\bibinfo{publisher}{Wiley}, \bibinfo{address}{New York},
  \bibinfo{year}{1975}).

\bibitem[{\citenamefont{Ichimaru}(1992)}]{faradayeffectbooks}
\bibinfo{author}{\bibfnamefont{S.}~\bibnamefont{Ichimaru}},
  \emph{\bibinfo{title}{Statistical Plasma Physics}}
  (\bibinfo{publisher}{Addison Wesley}, \bibinfo{address}{Massachussetts},
  \bibinfo{year}{1992}).

\bibitem[{\citenamefont{Verbiest et~al.}(1999)\citenamefont{Verbiest,
  Karunanen, and Persoons}}]{verbiest}
\bibinfo{author}{\bibfnamefont{T.}~\bibnamefont{Verbiest}},
  \bibinfo{author}{\bibfnamefont{M.}~\bibnamefont{Karunanen}},
  \bibnamefont{and} \bibinfo{author}{\bibfnamefont{A.}~\bibnamefont{Persoons}},
  \bibinfo{journal}{Phys. Rev. Lett.} \textbf{\bibinfo{volume}{82}},
  \bibinfo{pages}{3601} (\bibinfo{year}{1999}).

\bibitem[{\citenamefont{Charney}(1979)}]{charney}
\bibinfo{author}{\bibfnamefont{E.}~\bibnamefont{Charney}},
  \emph{\bibinfo{title}{The Molecular Basis of Optical Activity}}
  (\bibinfo{publisher}{Wiley}, \bibinfo{address}{New York},
  \bibinfo{year}{1979}).

\bibitem[{\citenamefont{Mohanty et~al.}(1998)\citenamefont{Mohanty, Nieves, and
  Pal}}]{mnp:nupip}
\bibinfo{author}{\bibfnamefont{S.}~\bibnamefont{Mohanty}},
  \bibinfo{author}{\bibfnamefont{J.~F.} \bibnamefont{Nieves}},
  \bibnamefont{and} \bibinfo{author}{\bibfnamefont{P.~B.} \bibnamefont{Pal}},
  \bibinfo{journal}{Phys. Rev. D} \textbf{\bibinfo{volume}{58}},
  \bibinfo{pages}{093007} (\bibinfo{year}{1998}).

\bibitem[{\citenamefont{Nieves and Pal}(1989)}]{np:pip}
\bibinfo{author}{\bibfnamefont{J.~F.} \bibnamefont{Nieves}} \bibnamefont{and}
  \bibinfo{author}{\bibfnamefont{P.~B.} \bibnamefont{Pal}},
  \bibinfo{journal}{Phys. Rev. D} \textbf{\bibinfo{volume}{39}},
  \bibinfo{pages}{652} (\bibinfo{year}{1989}).

\bibitem[{\citenamefont{Smith et~al.}(2000)}]{nsfpaper}
\bibinfo{author}{\bibfnamefont{D.~R.} \bibnamefont{Smith}}
  \bibnamefont{et~al.}, \bibinfo{journal}{Phys. Rev. Lett.}
  \textbf{\bibinfo{volume}{84}}, \bibinfo{pages}{4184} (\bibinfo{year}{2000}).

\bibitem[{\citenamefont{Condon}(1937)}]{condon}
\bibinfo{author}{\bibfnamefont{E.~U.} \bibnamefont{Condon}},
  \bibinfo{journal}{Rev. Mod. Phys.} \textbf{\bibinfo{volume}{9}},
  \bibinfo{pages}{432} (\bibinfo{year}{1937}).

\bibitem[{\citenamefont{Drude}(1959)}]{drude}
\bibinfo{author}{\bibfnamefont{P.}~\bibnamefont{Drude}},
  \emph{\bibinfo{title}{The Theory of Optics}} (\bibinfo{publisher}{Dover},
  \bibinfo{address}{New York}, \bibinfo{year}{1959}).

\bibitem[{\citenamefont{Post}(1962)}]{post}
\bibinfo{author}{\bibfnamefont{E.~J.} \bibnamefont{Post}},
  \emph{\bibinfo{title}{Fundamental Structure of Electromagnetics}}
  (\bibinfo{publisher}{North Holland}, \bibinfo{address}{Amsterdam},
  \bibinfo{year}{1962}).

\bibitem[{\citenamefont{Lakhtakia et~al.}(1986)\citenamefont{Lakhtakia,
  Varadan, and Varadan}}]{varadan}
\bibinfo{author}{\bibfnamefont{A.}~\bibnamefont{Lakhtakia}},
  \bibinfo{author}{\bibfnamefont{V.~K.} \bibnamefont{Varadan}},
  \bibnamefont{and} \bibinfo{author}{\bibfnamefont{V.~V.}
  \bibnamefont{Varadan}}, \emph{\bibinfo{title}{Time-harmonic Electromagnetic
  Fields in Chiral Media}} (\bibinfo{publisher}{Springer},
  \bibinfo{address}{New York}, \bibinfo{year}{1986}).

\bibitem[{\citenamefont{Kong}(1986)}]{kong1}
\bibinfo{author}{\bibfnamefont{J.}~\bibnamefont{Kong}},
  \emph{\bibinfo{title}{Electromagnetic Wave Theory}}
  (\bibinfo{publisher}{Wiley}, \bibinfo{address}{New York},
  \bibinfo{year}{1986}).

\bibitem[{\citenamefont{Kong}(1972)}]{kong2}
\bibinfo{author}{\bibfnamefont{J.~A.} \bibnamefont{Kong}},
  \bibinfo{journal}{Proc. IEEE} \textbf{\bibinfo{volume}{60}},
  \bibinfo{pages}{1036} (\bibinfo{year}{1972}).

\bibitem[{\citenamefont{Krowne}(1984)}]{krowne}
\bibinfo{author}{\bibfnamefont{C.~M.} \bibnamefont{Krowne}},
  \bibinfo{journal}{IEEE Trans. Antennas Propag.}
  \textbf{\bibinfo{volume}{32}}, \bibinfo{pages}{1224} (\bibinfo{year}{1984}).

\bibitem[{\citenamefont{Monzon}(1990)}]{monzon}
\bibinfo{author}{\bibfnamefont{J.~C.} \bibnamefont{Monzon}},
  \bibinfo{journal}{IEEE Trans. Antennas Propag.}
  \textbf{\bibinfo{volume}{38}}, \bibinfo{pages}{227} (\bibinfo{year}{1990}).

\bibitem[{\citenamefont{Hillion}(1993)}]{hillion3}
\bibinfo{author}{\bibfnamefont{P.}~\bibnamefont{Hillion}},
  \bibinfo{journal}{Phys. Rev. E} \textbf{\bibinfo{volume}{48}},
  \bibinfo{pages}{3060} (\bibinfo{year}{1993}).

\bibitem[{\citenamefont{Nieves and Pal}(1994)}]{np:thirdconst}
\bibinfo{author}{\bibfnamefont{J.~F.} \bibnamefont{Nieves}} \bibnamefont{and}
  \bibinfo{author}{\bibfnamefont{P.~B.} \bibnamefont{Pal}},
  \bibinfo{journal}{Am. J. Phys.} \textbf{\bibinfo{volume}{62}},
  \bibinfo{pages}{207} (\bibinfo{year}{1994}).

\bibitem[{\citenamefont{Hehl et~al.}(2002)\citenamefont{Hehl, Obukhov, and
  Rubilar}}]{Hehl:2002hr}
\bibinfo{author}{\bibfnamefont{F.~W.} \bibnamefont{Hehl}},
  \bibinfo{author}{\bibfnamefont{Y.~N.} \bibnamefont{Obukhov}},
  \bibnamefont{and} \bibinfo{author}{\bibfnamefont{G.~F.}
  \bibnamefont{Rubilar}}, \bibinfo{journal}{Int. J. Mod. Phys.}
  \textbf{\bibinfo{volume}{A17}}, \bibinfo{pages}{2695} (\bibinfo{year}{2002}),
  \eprint{gr-qc/0203105}.

\bibitem[{\citenamefont{Balakin and Lemos}(2002)}]{Balakin:2002mc}
\bibinfo{author}{\bibfnamefont{A.~B.} \bibnamefont{Balakin}} \bibnamefont{and}
  \bibinfo{author}{\bibfnamefont{J.~P.~S.} \bibnamefont{Lemos}},
  \bibinfo{journal}{Class. Quant. Grav.} \textbf{\bibinfo{volume}{19}},
  \bibinfo{pages}{4897} (\bibinfo{year}{2002}), \eprint{gr-qc/0209062}.

\end{thebibliography}
\end{document}